\documentstyle[aps,multicol,epsfig]{revtex}
\newcommand{\beq}{\begin{equation}}
\newcommand{\eeq}{\end{equation}}
\begin{document}
\title{Separation of dissipation from diffusion}
\author{Tong Zhou\\
Department of Physics,
University of California, Santa Barbara, CA 93106}
\date{\today}
\maketitle

\begin{abstract}

We study velocity correlations induced by diffusion and dissipation 
in a simple dissipative dynamical system.  We observe that diffusion, 
as a result of time reversible microscopic processes, leads to 
correlations with different spatial parity from those caused 
by dissipation, consisting of time irreversible microscopic processes.
The velocity correlations observed conflict with the ``molecular chaos''
assumption of the Boltzmann equation.  To account for the apparent
correlation structures in our simple model system,
we propose that they indicate geometric distortion
of the phase space.


\end{abstract}

\begin{multicols}{2}
\narrowtext

The second law of thermodynamics states that a system left to 
itself will evolve towards a state with maximum entropy, 
i.e.\ maximum randomness.  We are familiar with 
order due to interaction 
where, for example for systems in equilibrium at certain
temperature, different strength of interaction among 
microscopic constituents leads to states with different degrees
of order---solid, liquid, or gas.  In this letter
we study a different type
of order, order purely induced by diffusion and dissipation.

Nonequilibrium phenomena have long attracted much research 
interest.  The Boltzmann equation is probably the most celebrated
theory, yet its crucial assumption of ``molecular chaos'' needs
scrutiny before it can be applied to specific systems, 
see for example\cite{AGH}.
Linear response theory\cite{KTH} 
also has wide applications in condensed matter physics.
Some of recent studies include derivation of Ohm's Law for
the periodic Lorentz gas\cite{CELS}, the connection between
the smoothness of distribution function and the number of 
degrees of freedom\cite{BMG}, finite thermal
conductivity in 1d lattices\cite{GLPV} and directed current in
a 1d system\cite{FYZ}, on nonequilibrium
phase transition\cite{Arndt}, and on violation of the 
fluctuation-dissipation equality\cite{CDK}, among others.
However many theoretical works done on nonequilibrium phenomena, 
to different extent, assume certain mathematical
forms as the starting point of their investigations, which, 
though making the problems at hand mathematically attackable, may
miss generic properties of real physical systems because of
the limitation of the forms being used. 
On the other hand, with the development of modern computer
technology, numerical experiments have become a very powerful 
tool.  Here we carry out our investigation
by applying extensive numerical studies on a simple model system,
and try to form an understanding based on these results,
without introducing any {\it ad hoc} mathematical 
functional forms to characterize the system.

\begin{figure}
\centerline{\epsfxsize=9.6cm \epsfysize=0.65cm \epsfbox{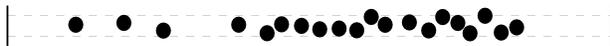}}
\vspace{0.2in}
\caption{A snapshot of the thin pipe system.  The periodic side walls are
indicated by dashed lines.  The coordinate system is set up so that the 
$x$-axis is along the pipe.}
\label{fg:pic}
\end{figure}

We study a two-dimensional system of identical non-interacting
hard disks confined 
in a thin pipe (Fig.~\ref{fg:pic}).  The width of the pipe is set so 
that two disks cannot pass each other.  Thus the motion of the disks 
is two dimensional, while at the same time their sequence is preserved.  
The two side walls are periodic and the two end walls are energy sources
kept at the same temperature.  For more details, see 
\cite{Zhou1,Zhou2}.  We use the simplest collision model---after a 
collision between two disks, 
the normal relative velocity changes sign, and decreases 
by a factor of the restitution coefficient $r$, with $0\le r\le 1$. 
In the collision, the other components of the velocities are 
unchanged.  Thus the total momentum is conserved while a portion
of kinetic energy is lost when $r<1$.
Simulations are for $80$ disks and the average spacing between 
two neighboring disks are much larger than their radii.  
Statistical analysis 
is done for the steady state.  Different versions of one-dimensional 
diffusion models are used in\cite{GLPV,FYZ,Arndt}.  However, they
rely on preset diffusion equations as their starting point, while 
we are attempting to find out the physical rules which will lead to
those equations.

Even with this simple model, we nevertheless 
limit ourselves to situations
close to equilibrium.  Extensive works have been done on such systems
where linear deviations from equilibrium are generally observed
and studied\cite{KTH,CELS,BMG,EvaM}.  
In our system, the details of the setting are so clearly 
defined that we may attempt to obtain a thorough understanding of
the linearity, which then may point to the direction for further 
study on situations farther away from equilibrium.

One of the cornerstones of classical kinetics 
theories, Liouville's theorem states that the phase space volume is a 
conserved quantity\cite{EvaM,LifP,Marsden}.  The importance
of underlying geometry of the phase space is especially stressed 
in\cite{TMK}.  It is long recognized that 
energy dissipation makes phase space tend to shrink\cite{Zhou1}.  
We would like to 
investigate the effects of this deviation from Liouville's theorem
on the dynamics.  The set up of the system is to create a very small, 
yet uniform dissipation and look for its first order effects.  To 
stay away from disturbing the internal dynamics, we put energy sources 
only at the end walls.  However, in this way we cannot have energy
dissipation without energy diffusion.  There are two processes going
on in the system.  First, energy dissipation due to inelastic collisions,
which to the first order is uniform.
Second, energy diffusion, meaning energy conduction from the end walls 
into the system, which at a certain location is proportional to its 
distance to the center---at the center there is no net diffusion in
either direction due to symmetry.  Energy conduction also exists
in a similar system with elastic collisions between disks
but two end walls at different temperature, which we will treat later.
The coexistence of
diffusion and dissipation is a general feature in dissipative systems.
We will demonstrate that they can be separated due to different 
spatial parities they exhibit in our system.

We concentrate on correlations among $x$-components, i.e. along the pipe, 
of velocities as an indicator of internal orders.  
For a system at equilibrium, even when there are interactions among its
constituents, their velocities obey Gaussian distribution and thus are 
uncorrelated.  The influence of a particle's motion on other particles' 
velocities is an indicator of order, and we quantify this order  with velocity
correlations.  Specifically, among other things,
we look at $\langle u_i|v_j\rangle$, 
meaning the conditional time average of the $x$-component of velocity 
of the $i$-th disk when $v_j$ is given, where $v_j\equiv u_{j+1}-u_j$.  Of
course, $\langle u_i\rangle$ by itself vanishes because in the
statistical steady state there is no average drift in disks' positions.

A somewhat more simplified quantity is the collision average\cite{Zhou2},
denoted by $\langle u_i\rangle_j$, which is the average of $u_i$ when the 
$j$-th disk is colliding with the $(j+1)$-st disk.  For a system in 
equilibrium, this quantity vanishes when $i$ is neither $j$ nor $j+1$.
However, in our system, it does not vanish, but exhibits regular and 
interesting patterns.  We can plot $\langle u_i\rangle_j$ as a function of
$i$ for several different $j$'s---the resulting curves are confusing.  But
if we plot the odd and even parity components of these curves, they are easily
identified as dissipation and diffusion, respectively.

\begin{figure}
\centerline{\epsfxsize=9cm \epsfbox{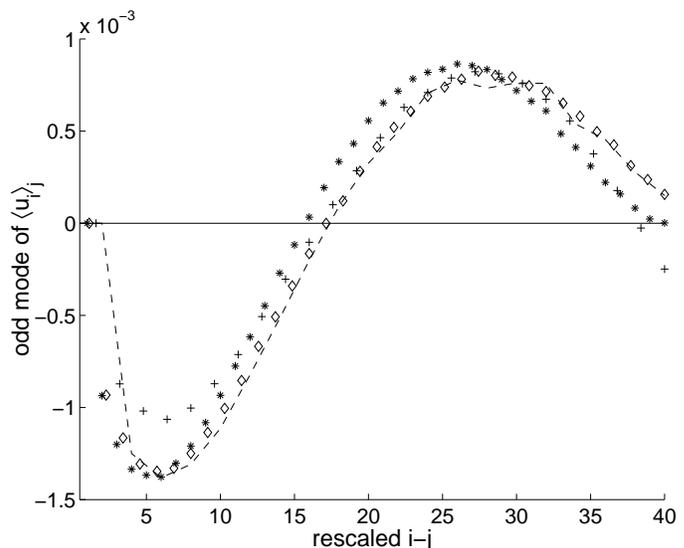}}
\caption{The odd parity component of $\langle u_i\rangle_j$.  $*$ is for
$j=40$, $\diamond$ for $j=35$ and $+$ for $j=25$.  The horizontal axis is for
$i-j$ rescaled so that the closest boundary to $j$ is shown at $40$.
Only the positive half of $i-j$ is shown.  The system has $80$ disks, 
$r=0.995$ and the two end walls are set to the same temperature.
The vertical axis is without
rescaling, demonstrating a uniform correlation structure from uniform
dissipation.  The dashed line is the corresponding rescaled curve for a system
of $40$ disks and $r=0.99$.}
\label{fg:odd}
\end{figure}

\begin{figure}
\centerline{\epsfxsize=9cm \epsfbox{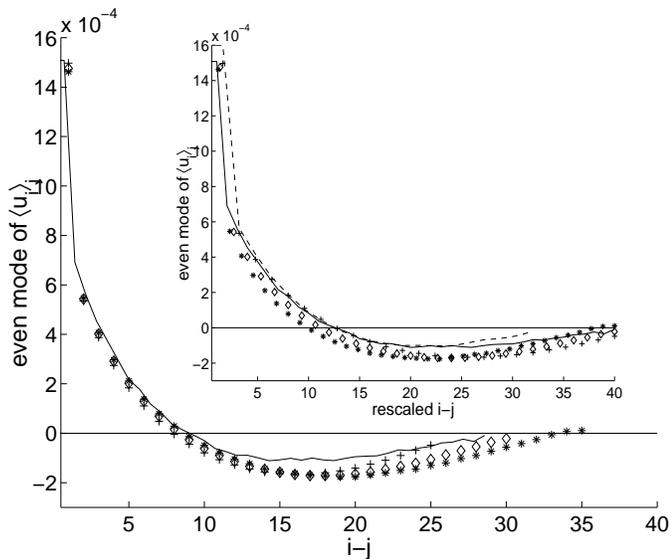}}
\caption{The even parity component of $\langle u_i\rangle_j$.  $*$ is for
$j=35$, $\diamond$ for $j=30$ and $+$ for $j=25$.  The horizontal axis is for
$i-j$ without rescaling.
Only the positive half of $i-j$ is shown.  For $j=35$, the vertical
coordination plotted is 
the real number divided by $5$ for the energy flux at $j=35$ supplies 
dissipation caused by $5$ disks; for $j=30$, it is 
the real number divided by $10$; for $j=25$, it is the real number divided
by $15$.  These three curves overlap, 
showing that the even mode is proportional to the distance between $j$ and 
the center.
The solid line is the corresponding rescaled curve for an elastic system
with only energy diffusion; its $i-j$ value is rescaled.
The inset is a similar plot, with rescaled $i-j$ as the horizontal axis, 
also shown as the dashed line is a corresponding curve from a elastic
system with $40$ disks.
}
\label{fg:even}
\end{figure}

The odd parity component, i.e., $\langle u_i-u_{2j+1-i}\rangle_j$ is for
dissipation---they overlap for different value of $j$'s (Fig.~\ref{fg:odd}), 
a clear signature 
for uniform dissipation.  Apparently, dissipation due to collision between
two neighboring disks has effects on the motions, not just of several 
disks close to them, but of all the disks in the system.
The even parity component, i.e., $\langle u_i+u_{2j+1-i}\rangle_j$ is for
diffusion---they are proportional to the distance between $j$ and the 
center $j=40$ (Fig.~\ref{fg:even}), a signature that they are caused
by energy conduction which is also proportional to this distance.

For a further test, we run simulations for a 
system with similar set up, but with elastic collisions and the two end walls
are kept at different temperatures so that no dissipation, only diffusion
occurs.  The odd parity component of the curves vanishes.  
And $\langle u_{i-j}\rangle_j$ is
independent of $j$, corresponding to uniform diffusion.  Also $\langle u_i
\rangle_j$ can be rescaled to overlap with the even parity components from the
dissipative system (Fig.~\ref{fg:even}).

Heuristically, we can explain the different spatial parities by 
pointing out that dissipation consists of
time irreversible microscopic processes
while diffusion consists of only time reversible microscopic processes.  
In the system with no dissipation but only constant diffusion, when the $j$-th
disk is colliding with the $(j+1)$-st disk, let us consider the following two
operations: time reverse and a spatial parity operation that reverses the 
direction of $x$-axis and interchanges 
$i$-th disk with $(2j+1-i)$-th disk.  These two operations leave
the energy flux unchanged and connect two equally probable microscopic states.
Because $u_i$ in one state is the same as $u_{2j+1-i}$ in the other state, 
we see the correlation curve from diffusion has even spatial parity.
The odd parity for the correlation curve from dissipation is intuitively
reasonable, given dissipation causes the shrinkage of phase space volume.

The correlation curves from diffusion and dissipation have following 
properties.
They are independent of the width of the pipe.  The diffusion curves 
are not sensitive as to
where the boundaries are---curves for different collision number $j$'s can 
overlap though the distances from $j$'s to the boundaries are different.  
However, they are not for local structures.  The inset of Fig.~\ref{fg:even}
shows the overlapping between diffusion curves for systems with $80$ 
disks (solid line) and $40$ disks (dashed line), 
only after suitable rescaling of $i-j$, which suggests
they are roughly functions of $(i-j)/N$.  This is true also for dissipative
curves, which only overlap with one another after rescaling of $i-j$ 
(Fig.~\ref{fg:odd}).  These observations suggest that the structures
from dissipation and diffusion are global.  

\begin{figure}
\centerline{\epsfxsize=9cm \epsfbox{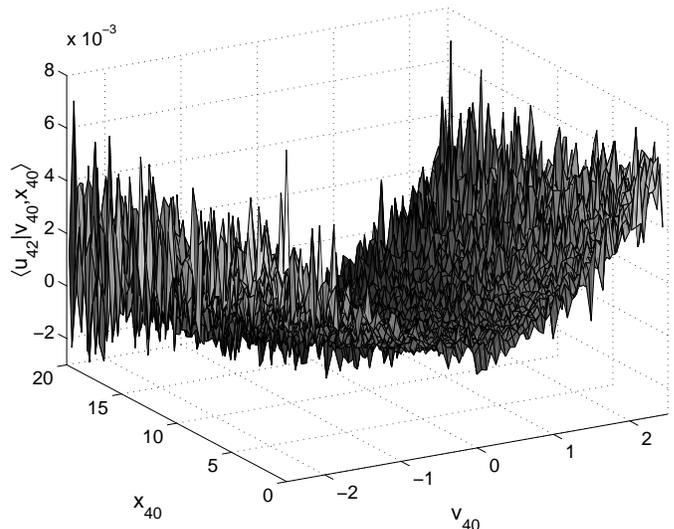}}
\caption{$\langle u_{42}|v_{40},x_{40}\rangle$.  Though noisy, the surface
plot shows a linear dependence on $x_{40}$ and a quadratic dependence on 
$v_{40}$.}
\label{fg:surfu}
\end{figure}

In the rest part of this letter, we describe our detailed investigation of
correlation curves for diffusion.  For this purpose, the simulations are for
systems with elastic collisions and different temperatures at two ends.
We look at the conditional time average of $u_i$ 
when $v_j$ and $x_j$ are given,
where $x_j$ is the spacing between $j$-th and $(j+1)$-st disks, and 
find (Fig.~\ref{fg:surfu}),
\beq
\langle u_i|v_j,x_j\rangle = c_{i-j}\left[(v_j^2-\langle v_j^2\rangle)
+f(x_j-\langle x_j\rangle)\right],
\eeq
where $f$ is a constant.  This suggests that the effects of $v_j$ on $u_i$ are
independent of those from $x_j$ and we can concentrate on just the velocities:
\beq
\langle u_i|v_j\rangle = c_{i-j}(v_j^2-\langle v_j^2\rangle).
\label{eq:str}
\eeq

It is not difficult to understand this relation for $i=j$.  Let us consider
$\langle u_j+u_{j+1}|v_j\rangle $ as a function of $v_j$.  It cannot be a
constant because 
$\langle (u_j+u_{j+1})v_j\rangle=\langle u_{j+1}^2-u_j^2\rangle <0$ 
while $\langle v_j\rangle=0$.  It cannot
have a linear term proportional to $v_j$
either because when $j$-th disk collides with $(j+1)$-st 
disk, $v_j$ changes sign while $u_j+u_{j+1}$ is unchanged due to momentum
conservation.  Then the simplest form $\langle u_j+u_{j+1}|v_j\rangle $ 
can have
is a constant times $v_j^2-\langle v_j^2\rangle $, as in (\ref{eq:str}).  
Though conforming to
quite general arguments,
the above observation is in contrast to the ``molecular chaos'' assumption
of the Boltzmann equation, the assumption that makes the equation time
irreversible.

It is counterintuitive that (\ref{eq:str}) does not vanish when $i$ is 
neither $j$ nor $j+1$.  The evidence shown in Fig.~\ref{fg:even} 
implies a global structure
yet depending on local inhomogeneity.  We propose an approach based on
geometric distortion of the phase space.  Let us think the phase space as
a Riemannian manifold with a metric $g$ for the usual $(p,q)$ coordinate
system\cite{TMK}.  
The statistical probability density $\rho$ is a scalar.  But the
probability of the system in the neighborhood of a point is proportional
to $\rho\sqrt{g}$.  One approach would be to view the phase space as 
a simple Euclidean manifold with a trivial $g$, then $\rho$ has to account
for the interesting but puzzling correlation structure.  Another approach,
which we prefer, is to set $\rho$ to a constant, and view the correlation
pattern as an indication of the distortion of the phase space and thus
a non-constant $g$.

\begin{figure}
\centerline{\epsfxsize=9cm \epsfbox{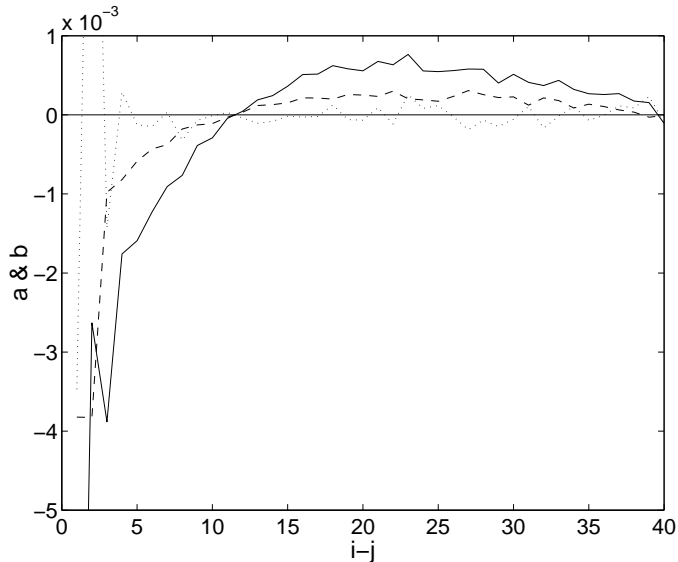}}
\caption{Coefficients $a(i-j)$ and $b(i-j)$ in (\ref{eq:retr}).  Solid line
is for $a$ and dashed line for $b$.  The first data point of $a$ is not shown,
which at about $-0.01$.  The dotted line is $a-2.5b$ and is close to $0$.}
\label{fg:abr}
\end{figure}

In this approach, because the distortion of the phase space, the variables
$u_i$'s are no longer statistical
independent and the corresponding coordinate system 
no longer orthogonal.  Let us assume a new set of
independent variables $u_i'$'s
deviates from the original set to the first order:
\beq
u_i' = u_i + \epsilon F(\Gamma),
\label{eq:trans}
\eeq
where $\epsilon$ is proportional to the small inhomogeneity, and $\Gamma$
denotes the set of $u_i$'s.  Because $\langle u_i'|v_j\rangle=0$
by definition, we have
$\langle u_i|v_j\rangle = -\epsilon \langle F(\Gamma)|v_j\rangle_0$.  The 
subscript $0$ in the last expression indicates that only the zeroth order is 
needed, which can be extracted from numerical data.  From the somewhat
noisy data, we find (\ref{eq:trans}) takes the form,
\beq
u_i' = u_i + \sum_{j=1}^{N-1}a_{i,j}v_j^2+\sum_{j=1}^{N-2}b_{i,j}v_{j-1}v_j+C,
\label{eq:retr}
\eeq
where $C$ is a constant that keeps $\langle u_i'\rangle=0.$  Let us for the
moment ignore boundaries and write the coefficients as $a(i-j)$ and $b(i-j)$
respectively.  $b(i-j)$ can be obtained by noticing:
\beq
\langle u_i|v_{j-1},v_j\rangle-\langle u_i|v_{j-1},-v_j\rangle = -2b(i-j)
v_{j-1}v_j.
\eeq
Then in turn $a(i-j)$ can be obtained by substituting $b(i-j)$ into
\begin{eqnarray}
\langle u_i|v_j\rangle & = &  \left[-a(i-j)-\frac{1}{4}a(i-(j-1))
-\frac{1}{4}a(i-(j+1))\right.\nonumber\\
&&\left.+\frac{1}{2}b(i-j)+\frac{1}{2}b(i-(j+1))\right]v_j^2.
\end{eqnarray}
We use numerical data from $\langle u_i|v_{40},v_{41}\rangle$ to extract 
values of $a(i-j)$ and $b(i-j)$ and plot the result in Fig.~\ref{fg:abr}.

We can assume the metric under the coordinate system $u_i'$ diagonal.
Then by the relation between two coordinate systems (\ref{eq:trans}),
we can obtain the metric
$g$ under the original coordinate system of $u_i$'s.  Because the metric
$g$ is connected to the strain tensor\cite{Lodge}, we see how the correlations
indicate the distortion of the phase space.

Because we study the behavior of the system very close to equilibrium,
we expect a smooth distribution function in the phase space, agreeing
with the approach of Tuckerman {\it et al.}\cite{TMK}.  
However, when the inhomogeneity
is stronger\cite{JaeNB}, there is likely to be a transition in the dynamics and
the distribution function becomes fractal as suggested by Hoover, {\it et al.}

In this letter, we present numerical results of velocity correlations in a
simple model of nonequilibrium system and propose to interpret the
observation as an indication 
of geometric distortion of the phase space.  However,
most theoretical reasonings in the letter
are heuristic arguments.  We did not attempt to
formulate a fundamental theoretical framework because we feel further data
collecting, and more importantly, different, more general setting of the
system should be tested before a better understanding can be reached.  A
simple extension would be the testing of a similar system, but with two
dimensional close-packed disks.

This work was supported by EPRI/DoD through the Program on Interactive
Complex Networks, and NSF Grant No. DMR-9813752, and by a grant from
the Keck Foundation.

\end{multicols}
\end{document}